\documentclass{article}
\usepackage{epsfig}

\begin{document}
\voffset 1.5cm

\begin{center}
{\LARGE \bf Microscopic study of neutron-rich Dysprosium isotopes}

\vspace{0.5cm}
Carlos E. Vargas\footnote{cavargas@uv.mx} $\dagger\&$, V\'ictor Vel\'azquez
$\&$, Sergio Lerma $\dagger$

\vspace{0.5cm}
$\dagger$ Facultad de F\'{\i}sica e Inteligencia Artificial, Universidad
Veracruzana. Sebasti\'an Camacho 5; Xalapa, Ver., 91000, M\'exico \\

\vspace{0.5cm}
$\&$ Facultad de Ciencias, Universidad Nacional Aut\'onoma de M\'exico, Apartado 
Postal 70-542, 04510 M\'exico D.F., M\'exico \\

\end{center}

\vskip 0.5cm
\date{\today}

\begin{abstract}

Microscopic studies in heavy nuclei are very scarce due to large valence 
spaces involved. This computational problem can be avoided by means of the use of 
symmetry based models. Ground-state, $\gamma$ and $\beta$-bands, 
and their B(E2) transition strengths in $^{160-168}$Dy isotopes, are 
studied in the framework of the pseudo-SU(3) model which includes the preserving symmetry 
$Q\cdot Q$ term and the symmetry-breaking Nilsson and pairing terms, systematically 
parametrized. Additionally, three rotor-like terms are considered whose free parameters,
fixed for all members of the chain are
used to fine tune the moment of inertia of rotational bands and the band-head of 
$\gamma$ and $\beta$-bands. The model succesfully describes
in a systematic way rotational features in these nuclei and allows 
to extrapolate toward the midshell nucleus $^{170}$Dy.
The results presented show that it is possible to study full chain of isotopes or isotones 
in the region with the present model.

\end{abstract}

\section{I. Introduction}
Rare-earth neutron-rich nuclei offer an opportunity to study singular aspects of nuclear structure around 
the midshell region such as new modes of excitation and collectivity, the role of the single-particle levels 
\cite{Wal99} and even possible modifications of shell structure. Nevertheless, despite the recent progress 
in experimental techniques, the rare-earth neutron-rich nuclei remains a region of the nuclear chart which 
has been seldom studied. For example, in heavy-ion induced multinucleon transfer reaction, the combination of 
stable projectiles and stable targets has unfavorable kinematic matching conditions leading to heavy 
neutron-rich nuclei, resulting in small cross sections that limits its use for populating neutron-rich nuclei 
\cite{Wu98}. In this region, nuclei have large numbers of valence protons and neutrons, and many low-lying 
collective nuclear properties are expected to appear there. For example, Er, Yb and Hf neutron-rich nuclei exhibit
a minimum in the $2^+$ energies and a maximum of deformation at N=104 \cite{Fir96}. Assuming standard spherical 
shell gaps, $^{170}$Dy has the larger number of valence particles of any nucleus lighter than the doubly magic 
lead, and hence it is expected to be the most collective nucleus in the region \cite{Boh75,Cas93,Zha01,Cas88}. 
Experimental works in the lighter dysprosium isotopes have shown that there is an enhancement of the collectivity in 
$^{164}$Dy, a relative flatness of the moment of inertia in $^{166}$Dy, and $^{168}$Dy shows again an apparent 
enhancement of deformation \cite{Asa99}. Nevertheless, to date the energies of the levels in $^{170}$Dy have not 
been measured, and hence its degree of collectivity is still unknown. In addition, the known saturation of 
B(E2) in the region \cite{Cas93,Zha01,Cas88} and the decrease in moment of inertia between N=98-100 in the gadolinium 
isotopes \cite{Jon04} open the following question of the collectivity in the region: is $^{170}$Dy indeed the most
collective of the region or is a little of the collectivity lost when we reach N=104.

Among the different chanels for producing the dysprosium neutron-rich nuclei, quasielastic 
two-neutron transfer reaction has been used to study the neutron-rich $^{166}$Dy 
\cite{Wu98}, two-proton pick-up reaction allowed to populate $^{168}$Dy \cite{Asa99,Xiu98}, 
and projectile fragmentation of $^{208}$Pb was used \cite{Pod00,Cam02} to synthetize the midshell 
$^{170}$Dy. 

From the theoretical side, the large valence spaces associated with heavy-nuclei have implied 
a slow progress, limiting the number of models capable of dealing with $^{170}$Dy. The earlier 
studies of this nucleus have used the relativistic mean-field calculations \cite{Lal96}, 
the Strutinsky shell correction \cite{Lon97,Naz90}, the generator co-ordinate method \cite{Ner87}, 
and the Monte Carlo shell-model \cite{Dea93}. More recently, P. H. Regan {\it et al.} 
\cite{Reg02,Rat03} used the cranked shell-model with nonaxial deformed Woods-Saxon potential, and predicted 
a highly deformed $^{170}$Dy nucleus with a pure axial symmetric-shape deformation along the Yrast line
and a $K^\pi$ = $6^+$ isomeric state at an approximated energy of 1.2 MeV.

A shell model description \cite{Cau05} of heavy nuclei requires further assumptions 
that include a systematic and proper truncation of the Hilbert space \cite{Stu02}. The 
symmetry based SU(3) shell model \cite{Ell58,Ell58b} has been successfully applied in light nuclei, 
where a harmonic oscillator mean-field and a residual quadrupole-quadrupole  
interaction can be used to describe dominant features of the nuclear spectra. 
However, the strong spin-orbit interaction renders the SU(3) model useless in 
heavier nuclei, while at the same time pseudo-spin emerges as a good symmetry \cite{Hec69,Ari69}. 
The origin of this symmetry has been traced back to the relativistic mean field for
heavy-nuclei \cite{Gin97,Blo95}, and the success of the pseudo-SU(3) model \cite{Rat73} 
lies on the consistency of this symmetry. On other hand, the backbending phenomenon
in $^{154-164}$Dy chain has been studied with the projected shell model \cite{Vel99}.

The first applications of the pseudo-SU(3) model considered pseudo-spin as a dynamical 
symmetry \cite{Dra82,Dra84,Cas87}. With the development of a computer code came a technical 
breakthrough that enabled mixed-representation calculations for calculating reduced 
matrix elements of physical operators (i.e. the symmetry-breaking Nilsson 
single-particle energies or pairing correlations) between different SU(3) irreps 
\cite{Bah94}. This enabling technology allows fully microscopic studies of 
energy levels belonging to normal parity bands and transitions 
\cite{Beu00,Hir02,Var04,Dra04,Hir06,Var02}. 

The study of abnormal parity bands of rare earth nuclei, in addition of those with normal 
parity, requires the explicit consideration of intruder levels, which may be performed 
using the quasi-SU(3) scheme, as it was pointed out in Refs. \cite{Zuk95,Mar97}, coupling the 
intruder states of the quasi-SU(3) with those of normal parity described by the pseudo-SU(3). 
In Reference \cite{Cau05} the authors report some calculations of the quadrupole properties 
of $^{48}$Cr and E2 transition probabilities of the 
Nd, Sm, Gd and Dy nuclei with 92 $\leq$ N $\leq$ 98, showing the value of this theoretical 
framework, because their predictions are very close to values ​​reported experimentally. In 
the $sd$-shell, the quasi-SU(3) scheme was applied to describe even-even, even-odd and 
odd-odd nuclei \cite{Var01c,Var01b}.

However, a calculation of rare-earth nuclei which considers the coupling of quasi- and 
pseudo SU(3)-schemes is impossible at this time due to computational challenges unresolved 
yet. The first step in the development of these codes would be the building of the basis, 
which will emerge of the coupling of normal with intruder states, mixing the occupation 
numbers associated to each subspace. In this case, the use of pseudo + quasi scheme 
(P and Q, respectively) would involve the following couplings

\begin{eqnarray}
(\tilde{\lambda_\pi},\tilde{\mu_\pi})_P \otimes (\lambda_\pi,\mu_\pi)_Q = (\lambda_\pi,\mu_\pi)_{P+Q}, \nonumber \\
(\tilde{\lambda_\nu},\tilde{\mu_\nu})_P \otimes (\lambda_\nu,\mu_\nu)_Q = (\lambda_\nu,\mu_\nu)_{P+Q}, \nonumber \\
(\lambda_\pi,\mu_\pi)_{P+Q} \otimes (\lambda_\nu,\mu_\nu)_{P+Q} = (\lambda,\mu)^{Total}_{P+Q}, \nonumber
\end{eqnarray}

\noindent where the first row is for protons, the second is for neutrons and the third 
shows the final coupling that might be used in the description of the whole nucleus and where 
it has not been considered the changes in occupation numbers. The second step would be the extension 
of the Hamiltonian to include the coupling with the particles in the intruder sector. These two 
major extensions would give us the ability of realize an comprehensive description of heavy 
nuclei, as for example the study of abnormal parity bands, and the reduction of huge effective 
charges used up to now. 

In order to illustrate this calculation, we can take as example the nucleus $^{160}$Dy. It has 66 
protons and 94-104 neutrons, and of these, 16 protons and 12 neutrons are in the last unfilled 
(open) shells. Assuming a deformation 
$\beta~\sim~0.25$, the deformed Nilsson single-particle levels of the active shells are filled 
from below \cite{Dra84,Cas87}. Ten protons are distributed in the $1g_{7/2}$ and $2d_{5/2}$ 
orbitals of the $\eta$ = 4 shell, and the remaining six occupy the $1h_{11/2}$ intruder orbital. 
Eight neutrons occupy the $2f_{7/2}$ and $1h_{9/2}$ orbitals of the $\eta$ = 5 shell and four are 
in $1i_{13/2}$ orbital. This calculation would consider a total of 28 nucleons in active shells, which 
increases the computational time by several orders of magnitude; so even considering the use of 
symmetries it would be necessary to modify the current codes to increase their efficiency. These 
studies are beyond the goals of this work and are left as future extensions of the model. 
Nevertheless, following the suggestions of the Refs. \cite{Cau05,Zuk95,Mar97}, it is possible to employ the 
quasi-SU(3) scheme to estimate the contributions of the intruder sector to the quadrupole moments ($Q_0$) 
and relate them to the E2 transition probability, eliminating the use of the extremely large 
effective charges.

In this Article, we present for the first time an application of the pseudo-SU(3) model
to the $^{160-168}$Dy chain, where the most relevant quadrupole-quadrupole, Nilsson 
single-particle and pairing terms are parametrized systematically and at the same time, 
a best fit of the parameters $a(K^2)$, $b(J^2)$ and $c(\tilde{C}_3)$ is done for this 
set of nuclei. The calculations for these known nuclei 
are extrapolated towards the neutron-rich nucleus $^{170}$Dy. Previous applications of the
model made use of several nucleus by nucleus fitting procedure of several parameters, not allowing 
a systematic study of the evolution of properties (for example, the modifications in the
moment of inertia of rotational bands) along chains of nuclei and making unreal the 
extrapolations predicting the spectroscopy 
of unmeasured nucleus. In the present approach, our goal is to study the evolution of
collectivity along the dysprosium isotopic chain (Z=66) starting at N = 94 and ending with
predictions for ground-state, $\gamma$ and $\beta$ bands in the N = 104 midshell $^{170}$Dy 
nucleus. To this end, the procedure we have followed is to employ a model with the 
Nilsson, Quadrupole-Quadrupole and Pairing terms of the Hamiltonian (\ref{eq:ham}) 
systematically parametrized \cite{Var00a} in function of the 
mass ($A$), whereas the parameters of the last three terms [$a$ $(K^2)$, $b$ $(J^2)$ and 
$c$ $(\tilde{C}_3)$] were determined by applying a best-fit procedure to the experimental 
data in all the chain.

In Section II a brief description of the pseudo-SU(3) classification scheme
and the Hamiltonian of the model are discussed. Theoretical energies of levels, comparision 
with data (when available) for ground-state, $\gamma$ and $\beta$ bands in $^{160}$Dy, $^{162}$Dy, 
$^{164}$Dy, $^{166}$Dy and $^{168}$Dy, predictions in $^{170}$Dy and the analysis of 
collectivity implied are presented in the Section III. Intra and inter-band B(E2) 
transitions are presented in Section IV, and finally, a brief conclusion 
is given in Section V.

\section{II. The model}

The selection of the many-body basis is the starting point for any shell-model
application. Many-particle states of $n_\alpha$ active nucleons
($\alpha = \pi, \nu$) in a given ($N$) normal
parity shell $\eta_\alpha^N$ are classified by the following group chain
\cite{Dra82,Dra84,Cas87,Var00a}:

\begin{eqnarray}
~ \{ 1^{n^{N}_\alpha} \} ~~~~~~~ \{ \tilde{f}_\alpha \} ~~~\{ f_\alpha
\} ~\gamma_\alpha ~~~ (\lambda_\alpha , \mu_\alpha ) ~~~ \tilde{S}_\alpha
~~ K_\alpha  \nonumber \\
U(\Omega^N_\alpha ) \supset U(\Omega^N_\alpha / 2 ) \times U(2) \supset
SU(3) \times SU(2) \supset \nonumber \\
\tilde{L}_\alpha  ~~~~~~~~~~~~~~~~~~~~~ J_\alpha ~~~~ \nonumber \\
SO(3) \times SU(2) \supset SU_J(2),
\label{eq:chains}
\end{eqnarray}

\noindent where above each group the quantum numbers that characterize its
irreducible representations (irreps) are given. $\gamma_\alpha$ and
$K_\alpha$ are multiplicity labels of the indicated reductions. The occupation 
numbers for protons are constant along the chain. In shell model applications, the 
dysprosium isotopes are considered to have 16 protons out of the Z = 50 inert core, 10 of these in
normal and 6 in abnormal $h_{11/2}$ parity levels. In Table \ref{occupations} the occupation 
numbers for neutrons assigned to each nucleus are presented.

\begin{table}
\begin{tabular}{ccccc}
&&&& \\ \hline \hline
     Nucleus        & $\epsilon_2$ & $n_\nu$ & $n_\nu^N$ & $n_\nu^A$ \\ \hline 
 $^{160}$Dy$_{94}$  &  0.250   &   12    &   8       &  4        \\
 $^{162}$Dy$_{96}$  &  0.258   &   14    &   8       &  6        \\
 $^{164}$Dy$_{98}$  &  0.267   &   16    &   10      &  6        \\
 $^{166}$Dy$_{100}$ &  0.267   &   18    &   12      &  6        \\
 $^{168}$Dy$_{102}$ &  0.275   &   20    &   12      &  8        \\
 $^{170}$Dy$_{104}$ &  0.267   &   22    &   14      &  8        \\ \hline \hline
\end{tabular}
\caption{Deformation ($\epsilon_2$) \cite{Mol95} and occupation numbers for neutrons ($n$). The 
superscript $N$ and $A$ indicate normal and abnormal parity levels, respectively.}
\label{occupations}
\end{table}

The first application of the pseudo-SU(3) model, where symmetry-breaking terms 
were included in the Hamiltonian, considered a Hilbert space composed of those normal parity 
states $ | \beta J M \rangle $ with the highest spatial symmetry, $\tilde{S}_{\pi,\nu}$~=~0 (for 
an even) and 1/2 (for an odd) number of protons or neutrons. That approach has allowed to describe 
in each nucleus typically three rotational bands in even-even \cite{Beu00,Hir02} and odd-mass \cite{Var00b} 
nuclei and intra- and inter-band \cite{Var01} B(E2) transition strengths. As has been the case 
for almost all studies with the model to date, nucleons in
abnormal parity orbital are considered to renormalize the dynamics that is described
using only nucleons in normal parity states. This limitation is reflected, for example,
by the use of very high effective charges to describe quadrupole electromagnetic 
transitions. While this is the most important limitation of the model and a very strong assumption, 
it has been shown to be a reasonable approach \cite{Var01b,Var98}.

In more recent studies \cite{Var04,Dra04,Hir06,Var02}, the extension of the Hilbert 
space to those lesser spatially symmetric states ($\tilde{S}_{\pi,\nu}$ = 1 and 3/2,  
for even and odd number of protons or neutrons, respectively) has shown that the model 
can describe up to eight excited bands (in each nucleus), intra- and inter-band B(E2) 
transition strengths between them, and to disscus the interplay between the collective and 
single-particle nature of the M1 excitations, the so-called scissors mode \cite{Var03}. 
A detailed analysis of the wave function content shows that the ground-state band is 
composed predominantly of $\tilde{S}_{\pi,\nu}$ = 0 and 1/2 states (for even and odd
number of $\pi$ and $\nu$, respectively) with a very small mixing of 
$\tilde{S}_{\pi,\nu}$ = 1 and 3/2 irreps. Nevertheless, in excited
bands (including the $\gamma$ and $\beta$ bands), the states with $\tilde{S}_{\pi,\nu}$ = 1 
and 3/2 have a very important contribution, and the correct description of many
spectroscopic properties requires a truncation scheme including the $\tilde{S}_{\pi,\nu}$ = 1 
states (in even-even nuclei). As we are dealing with even-even dysprosium isotopes and 
interested in the ground-state, $\gamma$ and $\beta$-bands, in the present work 
those states with $\tilde{S}_{\pi,\nu}$ = 0 and 1 are considered.

The Hamiltonian contains spherical Nilsson single-particle terms for the protons 
and neutrons ($H_{sp,\pi[\nu]}$), the quadrupole-quadrupole ($\tilde Q \cdot \tilde 
Q$) and pairing ($H_{pair,\pi[\nu]}$) interactions parametrized systematically 
\cite{Rin79,Duf96}, as well as three rotorlike terms ($K^2$, $J^2$ and $C_3$) that are 
diagonal in the SU(3) basis:

\begin{eqnarray}
    H & = & \sum_{\alpha=\pi,\nu} \{ H_{sp,\alpha} - G_\alpha ~H_{pair,\alpha}
	\} - \frac{1}{2}~  \chi~ \tilde Q \cdot \tilde Q \label{eq:ham} \\
      &   & + ~a~ K^2 + ~b~ J^2~ + ~c~ \tilde C_3. \nonumber
\end{eqnarray}

\noindent A detailed analysis of each term of this Hamiltonian and its systematic
parametrization can be found in Ref. \cite{Var00a}. The first row contains the basic 
components of any realistic Hamiltonian: the single-particle levels, pairing 
correlations and the quadrupole-quadrupole interaction, essential in the 
description of deformed nuclei. They have been widely studied in nuclear 
physics literature, allowing to fix their respective strengths by systematics 
($A$ dependent) \cite{Rin79,Duf96}, consequently they are not considered as free 
parameters of the model. The SU(3) mixing is due to the symmetry-breaking Nilsson 
single-particle and pairing terms.

The rotorlike terms in Hamiltonian (\ref{eq:ham}) are used to fine tune the spectra. Their 
three parameters $a$, $b$, and $c$ have been fixed following the prescriptions given in Ref. 
\cite{Var00a}, where a detailed analysis of each term can be found. Only these three terms 
are taken as free parameters of the model, and once their magnitude are determined by best 
fit, their values are kept constant for the full chain of isotopes. To clarify
the effect of these terms over the energies, we may take as example the $b J^2$ term. Its
effect is to give additional moment of inertia to the rotational bands, helping to diminish 
the energy of rotational states. When this term is added, the energy of the $J^\pi=2_{g.s.b.}^+$ 
states in the neutron-rich dysprosium isotopes reduces its value between 18 and 20 keV. 
For example, in $^{162}$Dy the energy of the $J^\pi=2_{g.s.b.}^+$ state changes from 94 
keV (when $b=0$) to 74 keV ($b=-3.2$ keV). In $^{164}$Dy it changes from 100 keV ($b=0$) 
to 80 keV ($b=-3.2$ keV). Those states with higher angular moment are also affected by this 
rotor term with larger changes. Using $^{164}$Dy as example, the state 
$J^\pi=8_{\gamma}^+$ changes from 1845 keV when $b=0$ to 1606 keV when $b=-3.2$
keV. It is important to mention that the rotor terms have influence on the level
spacings within the rotational bands, but the wave function holds practically constant
\cite{Var00a} within 0.001 \%.

\section{III. Low-lying energy spectra}
Using the basis and Hamiltonian presented in the previous section, we present the results 
for ground-state, $\gamma$ and $\beta$-bands in the $^{160-168}$Dy isotopes. In addition,
these results allow to extrapolate and predict energies and B(E2) transitions in $^{170}$Dy.

\begin{figure}
\epsfxsize=14.5cm
\centerline{\epsfbox{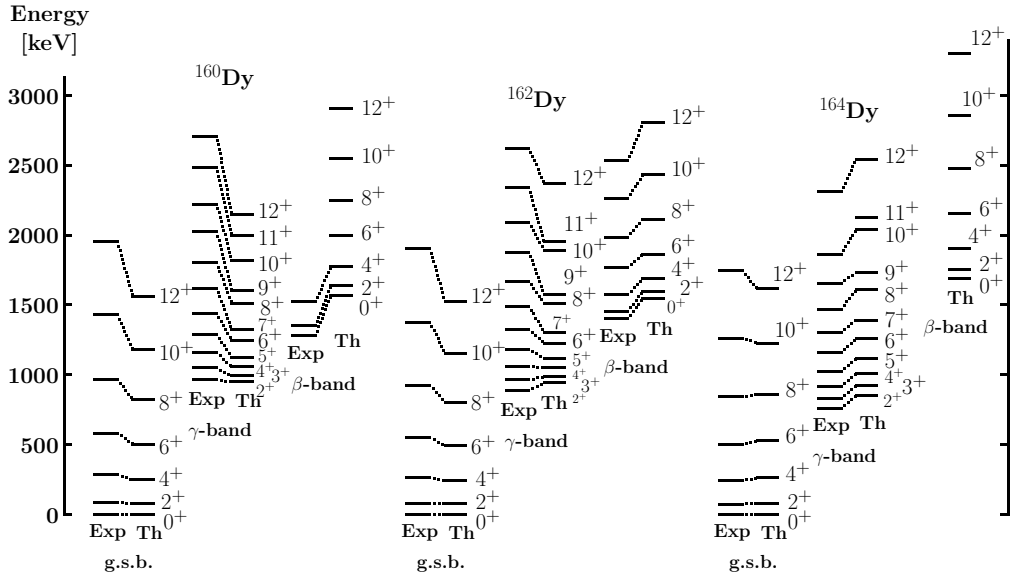}}
\caption{Experimental and theoretical energies (in keV) of ground-state, $\gamma$ and $\beta$-bands in 
$^{160-164}$Dy nuclei. The labels indicate the total angular momentum and parity of each level. 
Experimental data are plotted on the left-hand side of each column and theoretical ones on the
right-hand side. The correspondence between theoretical and experimental
levels is indicated by dotted-lines.}\vspace{-0.3cm}
\label{energies1}
\end{figure}

\begin{figure}
\epsfxsize=14.5cm
\centerline{\epsfbox{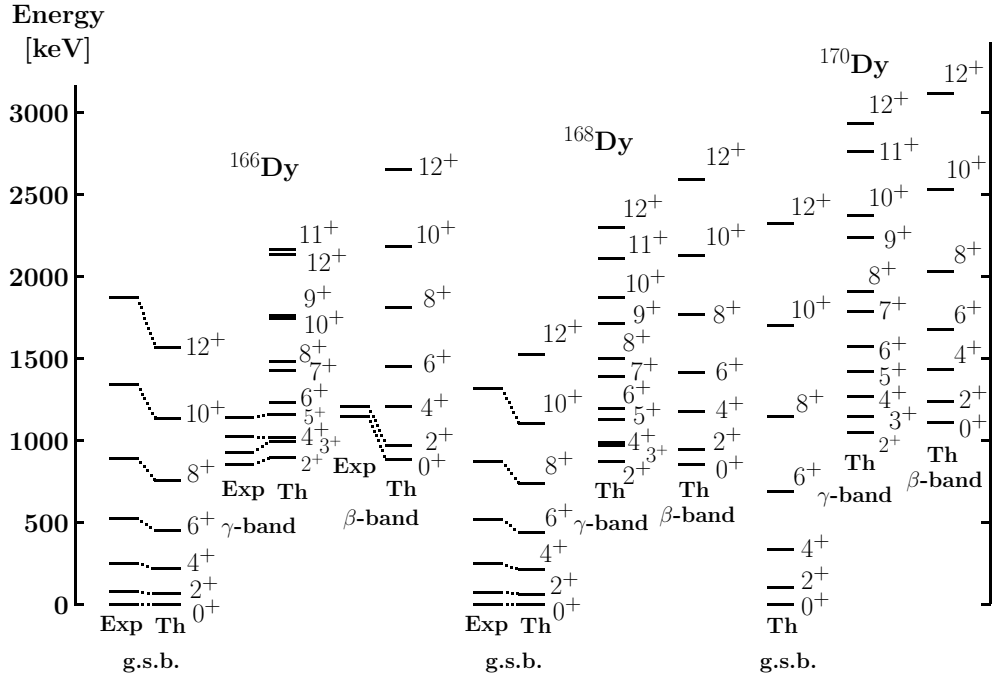}}
\caption{Experimental and theoretical energies (in keV) in $^{166-170}$Dy nuclei. 
The labels are the same as in Figure \ref{energies1}.}\vspace{-0.3cm}
\label{energies2}
\end{figure}

Since the value of the free parameters of the Hamiltonian (\ref{eq:ham}) have a very 
strong influence on the energies, their parametrization is very important. They are 
therefore determined by applying best fit to the experimental data for the five even-even 
isotopes $^{160-168}$Dy. The states from $J^\pi = 0^+$ to $6^+$ belonging to ground,
$\gamma$ and $\beta$ bands were used in the best fit procedure. Higher angular 
moments presented in the Figures \ref{energies1} and \ref{energies2} are not used in the procedure.
The values used in the present work are $a$ = 20, $b$ = -3.2 and $c$ = 0.033 (all in [keV]). 
The corresponding root-mean-square to the experimental data (from $J^\pi = 0^+$ to $6^+$) 
is 111 keV. That value allows one to estimate how good the fit is. It should be pointed out 
that the absolute values of the present free parameters are significantly smaller than those 
considered in previous works \cite{Var02,Var04,Dra04} in the region, where they got variations from 
positive to negative values when going from one nucleus to the next. This fact comes from the 
strategy followed in the present work, where a single value of each parameter has been used along
the chain, while previous applications of the model to the region used a different parametrization
for each nucleus, finding variations in the parameter from one nucleus to another, 
some times of one order of magnitude, or even in the sign. This implies that a greater 
rms than in previous applications of the model is expected; this is the cost 
when we want to make the model more predictive. On the other hand,
the benefits are that the more physically relevant terms of the Hamiltonian (first row
of eq. \ref{eq:ham}) are determining completely the structure of spectra, since
the contribution of the free parameters is very small, and that we can extrapolate and predict 
the energies of states in unknown nuclei. 

In the present application of the model, we present results for the energies of levels and
B(E2) transitions in the dysprosium (Z = 66) isotopic chain, including the stable $^{160}$Dy$_{94}$, 
$^{162}$Dy$_{96}$ and $^{164}$Dy$_{98}$, and the unstable $^{166}$Dy$_{100}$ and 
$^{168}$Dy$_{102}$ isotopes. Finally, the extrapolation allows us to predict the energies 
in the $^{170}$Dy$_{104}$ nucleus.

Figures \ref{energies1} and \ref{energies2} show the experimental \cite{Asa99,Xiu98,Bal01,Cor08,Cor02,Shi94} and 
theoretical energies for ground-state, $\gamma$ and $\beta$-bands in the $^{160-164}$Dy and $^{166-170}$Dy isotopic 
chains, respectively. For the ground-state band, the model predicts smaller energies than the experimental 
data, except in $^{164}$Dy (Figure \ref{energies1}) where the model overestimates almost all energies. 
The pattern of decreasing energy observed in experimental data in $^{164}$Dy is not described by the model. 
The case of the $\gamma$ band is different, where the model reproduces
the decrease in energy of the band head in $^{164}$Dy and higher values in the other nuclei, 
corroborating the trend observed in experimental data (when available). As it happens with the 
ground-state band, theoretical $\gamma$ bands have an overestimated moment of inertia, 
except in $^{164}$Dy where the model overestimates almost all energies. 
For the theoretical predictions, a common feature in the isotopic chain studied here (except 
$^{164}$Dy) is that the model predicts a strong collectivity as is implied by the data.
The lack of moment of inertia observed in theoretical values for $^{164}$Dy is probably due 
to the existence of some local effect that enhance nuclear deformation in this nucleus \cite{Asa99}.
In $\beta$ bands the model agrees well with experimental values, with higher values of the $0^+$ band-head
in $^{160}$Dy and $^{162}$Dy isotopes, and lower value in $^{166}$Dy. The nucelus $^{164}$Dy 
has peculiar rotational features, as it has been pointed out in Ref. \cite{Leh99}. It has been 
actively discussed the interpretation of the $\beta$-band in this nucleus, because from the experimental 
side there is not $\beta$-band reported up to now.

The predictions for the nucleus $^{170}$Dy are presented in the last three columns of Figure 
\ref{energies2}. It is surprising that the energies of levels predicted for this nucleus are 
higher than many presented in Figures \ref{energies1} or \ref{energies2} for other nuclei,
for the respective states. This represents a first indication of the model that $^{170}$Dy is 
less deformed and collective than the lighter dysprosium isotopes studied here. The same tendency 
has been identified in $^{162-164}$Gd isotopes \cite{Jon04}, where the moment of inertia between 
$N = 98-100$ decrease. It should be pointed out that the absence of nucleons in the intruder 
sector could help to explain the loss of collectivity in $^{164}$Dy and $^{170}$Dy, but at this 
stage of calculations there is no explanation for this loss of collectivity, presenting a challenge 
for future works. In addition, the model describes the 
$K^\pi = 6^+$ isomeric state which develops a rotational band. Its band-head is predicted at 
an energy of 1048 keV, very close to that predicted in Ref. \cite{Reg02} at 1200 keV. Here the band 
has its origin in the rotation of the intrinsec $K^\pi$ = $6^+$ state with pseudo-spin zero.
Future calculations in other nuclei of the region with high collectivity and
results coming from different models will contribute to confirm or reject the tendency
discussed in the present work.

\section{IV. Intra and inter-band B(E2)s}
In Table \ref{be2} we show intra-band B(E2) transition strengths up to $J^\pi = 8^+$ in the
ground-state, $\gamma$ and $\beta$ bands. The values are very collective 
between 200 and 700 $e^2 b^2 \times 10^{-2}$. The experimental data are between parentheses. 
Unfortunately, there are no experimental measurements for almost all transitions in $\gamma$
and $\beta$ bands. 

The effective charges used in the electric quadrupole operator $Q_\mu$ \cite{Var00a} are 
$e_\pi = 2.3$ and $e_\nu=1.3$. These values are the same used in the pseudo-SU(3) studies up to 
now allowing to describe both intra- and inter-band B(E2)s. They are larger than those used in 
standard calculations of quadrupole transitions \cite{Rin79} due to the absence of nucleons 
in intruder levels, and they were not varied to fit any particular value.

\begin{table}
\begin{tabular}{ccccccc}\hline \hline
 & \multicolumn{6}{c}{B(E2) [$e^2b^2 \times 10^{-2}$]}\\
$J_{i,band}^{\pi} \rightarrow J_{f,band}^{\pi}$ & $^{160}$Dy & $^{162}$Dy & $^{164}$Dy & $^{166}$Dy & $^{168}$Dy & $^{170}$Dy \\ \hline
% ground state band transitions  J --> J + 2
$0^+_{g.s.}  \rightarrow 2^+_{g.s}$    & 591 (500) & 596 (535) & 664 (557) & 717 & 723 & 712 \\
$2^+_{g.s.}  \rightarrow 4^+_{g.s}$    & 302 (146) & 305 (151) & 342 (145) & 368 & 371 & 365 \\
$4^+_{g.s.}  \rightarrow 6^+_{g.s}$    & 265 (122) & 267 (157) & 302 (173) & 323 & 325 & 322 \\
$6^+_{g.s.}  \rightarrow 8^+_{g.s}$    & 249 (170) & 251 (181) & 287 (165) & 302 & 304 & 304 \\
$8^+_{g.s.}  \rightarrow 10^+_{g.s}$   & 239 (169) & 241 (183) & 279 (188) & 288 & 290 & 292 \\
$10^+_{g.s.} \rightarrow 12^+_{g.s}$   & 232 (160) & 234 (173) & 274 (189) & 275 & 277 & 278 \\\hline
% Beta band transitions J --> J + 2
$0^+_{\beta} \rightarrow 2^+_{\beta}$  & 298       & 269       & 638       & 424 & 616 & 600 \\
$2^+_{\beta} \rightarrow 4^+_{\beta}$  & 127       & 160       & 294       & 141 & 140 & 282 \\
$4^+_{\beta} \rightarrow 6^+_{\beta}$  & 138       & 173       & 240       &  68 &  71 & 275 \\
$6^+_{\beta} \rightarrow 8^+_{\beta}$  & 148       & 165       & 236       &  69 &  73 & 173 \\
$8^+_{\beta} \rightarrow 10^+_{\beta}$ & 205       & 200       & 160       & 144 & 150 & 233 \\
$10^+_{\beta}\rightarrow 12^+_{\beta}$ & 198       & 151       & 201       & 144 & 144 & 227 \\\hline
% Gamma band transitions J --> J + 1
$2^+_{\gamma}\rightarrow 3^+_{\gamma}$ & 289       & 292       & 329       & 194 & 181 & 237 \\
$3^+_{\gamma}\rightarrow 4^+_{\gamma}$ & 192       & 193       & 217       &  57 &  61 & 151 \\
$4^+_{\gamma}\rightarrow 5^+_{\gamma}$ & 133       & 134       & 153       &  99 &  94 &  77 \\
$5^+_{\gamma}\rightarrow 6^+_{\gamma}$ &  88       &  88       &  95       &  81 &  82 &  67 \\
$6^+_{\gamma}\rightarrow 7^+_{\gamma}$ &  75       &  77       &  82       &  48 &  49 &  32 \\
$7^+_{\gamma}\rightarrow 8^+_{\gamma}$ &  33       &  19       &  45       &  67 &  16 &  14 \\\hline
% Gamma band transitions J --> J + 2
$2^+_{\gamma}\rightarrow 4^+_{\gamma}$ & 122       & 123       & 139       & 102 &  99 &  84 \\
$3^+_{\gamma}\rightarrow 5^+_{\gamma}$ & 172       & 173       & 117       &  50 &  54 & 205 \\
$4^+_{\gamma}\rightarrow 6^+_{\gamma}$ & 186       & 184       & 214       & 135 & 134 & 236 \\
$5^+_{\gamma}\rightarrow 7^+_{\gamma}$ & 195       & 197       & 234       & 156 & 159 & 245 \\
$6^+_{\gamma}\rightarrow 8^+_{\gamma}$ & 130       &  96       & 223       & 134 & 123 & 173 \\
$7^+_{\gamma}\rightarrow 9^+_{\gamma}$ & 179       & 181       & 242       & 152 & 152 & 256 \\ \hline \hline
\end{tabular}
\caption{Theoretical B(E2;$J^+_i \rightarrow J^+_f$)  [given in $e^2b^2 \times 10^{-2}$]
intra-band transitions in $^{160-170}$Dy nuclei. Known experimental data
\cite{Bal01,Cor08,Cor02,Shi94} are shown between parentheses. Effective charges are $e_\pi$=2.3 and $e_\nu$=1.3.}
\label{be2}\end{table}

The calculation of the electric quadrupole moment and B(E2) transition strengths with realistic effective charges ($e_\pi = 1.4$ 
and $e_\nu=0.6$) can be done if the intruder sector is explicitly considered. As discussed in the introduction, even if  a 
complete treatment of this sector is beyond the scope of this work, it can be obtained a simple evaluation of its effect by 
considering the quasi-SU(3) symmetry of the intruder sector, as proposed in Refs. \cite{Zuk95,Mar97}. Having established for a 
given deformation the number of nucleons in the abnormal parity levels, their contribution to $Q_0$ is easily obtained by 
filling the orbits in the right panel of Fig. 34 in Ref. \cite{Cau05}. By summing this contribution and that of the normal 
parity sector, the B(E2) transitions are calculated with realistic values for effective charges $e_\pi$ and $e_\nu$. In Table 
\ref{q+p} we show B(E2) intra-band transition strengths between the $0^+$ and $2^+$ states of the ground-state band.

\begin{table}
\begin{tabular}{ccccccc}\hline \hline
 & \multicolumn{6}{c}{B(E2) [$e^2b^2 \times 10^{-2}$]}\\
$J_{i,band}^{\pi} \rightarrow J_{f,band}^{\pi}$ & $^{160}$Dy & $^{162}$Dy & $^{164}$Dy & $^{166}$Dy & $^{168}$Dy & $^{170}$Dy \\ \hline
$0^+_{g.s.b.}  \rightarrow  2^+_{g.s.b}$        &  565       &  626       &  653       & 673        & 739        & 741       \\ \hline \hline
\end{tabular}
\caption{Pseudo + quasi-SU(3) B(E2;$0^+_{g.s.b.} \rightarrow 2^+_{g.s.b.}$)  [given in $e^2b^2 \times 10^{-2}$]
intra-band transitions in $^{160-170}$Dy nuclei . Effective charges are $e_\pi = 1.4$ and $e_\nu=0.6$.}
\label{q+p}\end{table}

By comparing the B(E2) values in Table \ref{q+p} with those of Table \ref{be2}, it can be clearly seen that the values in 
both tables  are very similar. Therefore, the closeness of both results implies that the explicit consideration of the intruder 
sector allows using realistic values for the effective charges $e_\pi$ and $e_\nu$. In other words, it means that the huge 
effective charges used in Table \ref{be2} are adequately compensating the absence of the intruder sector, which was not 
explicitly considered in the pseudo-SU(3) scheme.

Table \ref{interband} reports the inter-band B(E2) strengths between states of ground-state and $\gamma$
bands. These values are significantly smaller than those shown in Table \ref{be2}, because the
wave functions of states belong to different bands have very different components. Nevertheless, there are
some B(E2)s with large values, which is the result of a strong overlap between the wave functions
of the states.

\begin{table}
\begin{tabular}{ccccccc}\hline \hline
 & \multicolumn{6}{c}{B(E2) [$e^2b^2 \times 10^{-2}$]}\\
$J_{i,band}^{\pi} \rightarrow J_{f,band}^{\pi}$ & $^{160}$Dy & $^{162}$Dy & $^{164}$Dy & $^{166}$Dy & $^{168}$Dy & $^{170}$Dy \\ \hline
$0^+_{g.s.b.}  \rightarrow  2^+_{\gamma}$       &  14.4      &  14.6      &  16.5      & 2.3        & 1.7        & 18.4       \\
$2^+_{g.s.b.}  \rightarrow  3^+_{\gamma}$       &  7.1       &  7.3       &  8.3       & 2.7        & 2.6        & 9.0        \\
$3^+_{\gamma}  \rightarrow  4^+_{g.s.b.}$       &  4.2       &  4.3       &  9.1       & 1.3        & 1.2        & 5.9        \\\hline \hline
\end{tabular}
\caption{Theoretical B(E2;$J^+_i \rightarrow J^+_f$)  [given in $e^2b^2 \times 10^{-2}$]
inter-band transitions in $^{160-170}$Dy nuclei. Effective charges are the same as in Table \ref{be2}.}
\label{interband}\end{table}

The total pseudo-spin content of the nuclear wave function is built through the coupling of the 
$\tilde{S}_\pi$ and $\tilde{S}_\nu$ components. As it has been the case in previous works with 
the model, the ground-state bands in the chain of isotopes $^{160-170}$Dy are composed
predominantly by $\tilde{S}=0$, with very small mixing of the $\tilde{S}=1$, varying from 0$\%$
in $^{164}$Dy to 19$\%$ in $^{160}$Dy. The $\gamma$-bands have larger components of $\tilde{S}=1$,
varying from 7$\%$ in $^{164}$Dy to 90$\%$ in $^{166}$Dy. Large components of $\tilde{S}=1$ are
also observed in $\beta$-bands. The pseudo-spin contents for each 
band are practically constant along all states of the band. These results show the importance of 
the $\tilde{S}=1$ contribution in the description of excited $\gamma$ and $\beta$-bands.

\section{V. Conclusions}

The pseudo-SU(3) shell model offers a quantitative microscopic description of
heavy deformed nuclei. For the first time, employing a systematically parametrized
Hamiltonian and the best fit of three parameters for a set of nuclei, the model 
has been used to study energies and B(E2) transitions of the chain of dysprosium isotopes
and to predict the excitation energies in the ground-state, $\gamma$ and $\beta$-bands in $^{170}$Dy. 
The model describes a maximum of collectivity in $^{168}$Dy, but it fails in $^{164}$Dy where no enhancement 
of collectivity is found. In $^{170}$Dy a lower degree of collectivity is found, corroborating 
the tendency found in $^{162-164}$Gd isotopes \cite{Jon04}, where the moment of inertia between 
$N = 98-100$ decrease. At the present, the explanation for this loss of collectivity in 
$^{164}$Dy and $^{170}$Dy is the absence of a dynamical treatment of the intruder states in the model, 
presenting a challenge for future works.

Intra- and inter-band B(E2) transition strengths were presented, showing very collective bands
and some degree of mixing between ground-state and $\gamma$ and $\beta$ bands. Likewise, it was 
shown that the explicit consideration of the intruder sector using the quasi-SU(3) symmetry
allows to employ realistic effective charges. Nevertheless, the large effective charges used in 
our calculations work adequately to compensate the absence of the intruder sector in our model, as it
was checked for the B(E2;$0_{g.s.b.}^+\rightarrow 2_{g.s.b.}^+$) strength, where the values 
obtained with the large effective charges are similar to those obtained with realistic effective charges 
when is considered the effect of the abnormal parity states within a quasi-SU(3) scheme. A more 
complete study of the rare earth neutron-rich region, exploiting the quasi-SU(3) symmetry to 
include dynamically the intruder states, would be a natural extension of the present work.

In addition, we have presented the prediction of the model for the $K^\pi = 6^+$ 
isomeric state in $^{170}$Dy, which is at an energy of 1048 KeV, lower that the value of 
Ref. \cite{Reg02}. Finally, the results presented confirm the adequacy 
of the model to predict properties of rare-earth exotic nuclei. Experimental 
information for the neutron-rich nuclei is highly desired to understand nuclear structure 
around the midshell region.

\section{Acknowledgments}

The authors are grateful to ICN-UNAM for the valuable bibliographical support. This work was 
supported in part by CONACyT (M\'exico). CV would like to thank to Dr. Fred Mason Lamb​ert the 
style corrections and to Dr. A. E. Stuchbery and Dr. O. Casta\~nos for their valuable suggestions
to the present work.


\begin{thebibliography}{99}

\bibitem{Wal99} P. M. Walker, G. D. Dracoulis, Nature (London) 399, 35 (1999).
\bibitem{Wu98} C. Y. Wu, M. W. Simon, D. Cline, G. A. Davis, A. O. Macchiavelli and K. Vetter, 
	Phys. Rev. C 57, 3466 (1998).
\bibitem{Fir96} Firestone R. B. and Shirley V. S. (ed) 1996 {\it Table of isotopes} 8th edn (New York: Wiley).
\bibitem{Boh75} Aage Bohr, Ben R. Mottelson, ``Nuclear Structure", World Scientific Publishing 
	Company (January 15, 1998), Vol. II: Nuclear Deformations, Singapore.
\bibitem{Cas93} R. F. Casten and N. V. Zamfir, Phys. Rev. Lett. 70, 402 (1993).
\bibitem{Zha01} Y. M. Zhao, A. Arima, and R. F. Casten, Phys. Rev. C 63, 067302 (2001).
\bibitem{Cas88} R. F. Casten, J. Phys. G: Nucl. Phys. 14, S71 (1988).
\bibitem{Asa99} M. Asai {\it et al.}, Phys. Rev. C 59, 3060 (1999).
\bibitem{Jon04} E. F. Jones {\it et al.}, J. Phys. G: Nucl. Part. Phys. 30, L43 (2004).
\bibitem{Xiu98} Xiuqin Lu {\it et al.}, Eur. Phys. J. A 2, 149 (1998).
\bibitem{Pod00} Zs. Podolyak {\it et al.}; Proc. 2nd Intern. 
	Conf. Fission and Properties of Neutron-Rich Nuclei, St Andrews, Scotland, June 28-July 3, 
	1999, Ed. by J. H. Hamilton, W. R. Phillips, H. K. Carter, World Scientific, Singapore, p.156 (2000).
\bibitem{Cam02} M. Caama\~no, Ph.D. thesis, U. of Surrey, 2002.
\bibitem{Lal96} G. A. Lalazissis, M. M. Sharma, and P. Ring, Nucl. Phys. A 597, 35 (1996).
\bibitem{Lon97} G. L. Long, Chin. Phys. Lett. 14, 340 (1997).
\bibitem{Naz90} W. Nazarewicz, M. A. Riley, and J. D. Garrett, Nucl. Phys. A 512, 61 (1990).
\bibitem{Ner87} B. Nerlo-Pomorska, Z. Phys. A 328, 11 (1987).
\bibitem{Dea93} D. J. Dean, S. E. Koonin, G. H. Lang, W. E. Ormand and P. B. Radha, Phys. Lett. 
	B 317, 275 (1993).
\bibitem{Reg02} P. H. Regan, F. R. Xu, P. M. Walker, M. Oi, A. K. Rath and P. D. Stevenson, 
	Phys. Rev. C 65, 037302 (2002).
\bibitem{Rat03} A. K. Rath, P. D. Stevenson, P. H. Regan, F. R. Xu, and P. M. Walker, Phys. Rev. C 68, 044315 (2003).
\bibitem{Cau05} E. Caurier, G. Martinez-Pinedo, F. Nowacki, A. Poves, A. P. Zuker, Rev. Mod. Phys. 
	77, 427 (2005).
\bibitem{Stu02} A. E. Stuchbery, Nucl. Phys. A 700, 83 (2002).
\bibitem{Ell58} J. P. Elliott, Proc. Roy. Soc. London Ser. A {\bf 245}, 128 (1958).
\bibitem{Ell58b} J. P. Elliott, Proc. Roy. Soc. London Ser. A {\bf 245}, 562 (1958).
\bibitem{Hec69} K. T. Hecht and A. Adler, Nucl. Phys. {\bf A 137}, 129 (1969).
\bibitem{Ari69} A. Arima, M. Harvey, and K. Shimizu, Phys. Lett. {\bf B 30}, 517 (1969).
\bibitem{Gin97} J. N. Ginocchio, Phys. Rev. Lett. 78, 436 (1997); 
\bibitem{Blo95} A. L. Blokhin, C. Bahri and J. P. Draayer, Phys. Rev. Lett. 74, 4149 (1995).
\bibitem{Rat73} R. D. Ratna Raju, J. P. Draayer and K. T. Hetch, Nucl. Phys. {\bf A 202}, 433 (1973).
\bibitem{Vel99} V. Velazquez, J. G. Hirsch, Y. Sun, M. W. Guidry, Nucl. Phys. A 653, 17 (1999).
\bibitem{Dra82} J. P. Draayer, {\it et. al.}, Nucl. Phys. A {\bf 381}, 1 (1982).
\bibitem{Dra84} J. P. Draayer, and K. J. Weeks, Ann. of Phys. {\bf 156}, 41 (1984).
\bibitem{Cas87} O. Casta\~nos, J. P. Draayer, and Y. Leschber, Ann. Phys. {\bf 180}, 290 (1987).
\bibitem{Bah94} C. Bahri and J.P. Draayer, Comput. Phys. Commun. {\bf 83}, 59 (1994).
\bibitem{Beu00} T. Beuschel, J. G. Hirsch, and J. P. Draayer, Phys. Rev. {\bf C 61}, 054307 (2000).
\bibitem{Hir02} J. G. Hirsch, G. Popa, C. E. Vargas, and J. P. Draayer, Heavy Ion Physics {\bf 16}, 291 (2002).
\bibitem{Var04} C. E. Vargas, and J. G. Hirsch, Phys. Rev. C 70, 064320 (2004).
\bibitem{Dra04} J. P. Draayer, G. Popa, J. G. Hirsch and C. E. Vargas, High Energy Phys. and Nucl. Phys. 28, 1297 (2004).
\bibitem{Hir06} J. G. Hirsch, G. Popa, S. R. Lesher, A. Aprahamian, C. E. Vargas, and J. P. Draayer, 
	Rev. Mex. Fis. S {\bf 52}, 69 (2006).
\bibitem{Var02} C. E. Vargas, J. G. Hirsch, and J. P. Draayer, Phys. Rev. C {\bf 66}, 064309 (2002).
\bibitem{Zuk95} A.P. Zuker, J. Retamosa, A. Poves and E. Caurier, Phys. Rev. C {\bf 52}, R1741 (1995).
\bibitem{Mar97} G. Mart\'inez-Pinedo, A. P. Zuker, A. Poves and E. Caurier, Phys. Rev. C {\bf 55}, 187 (1997).
\bibitem{Var01c} C. E. Vargas, J. G. Hirsch, and J. P. Draayer, Nucl. Phys. A 690, 409 (2002).
\bibitem{Var01b} C. E. Vargas, J. G. Hirsch, and J. P. Draayer, Nucl. Phys. A 697, 655 (2002).
\bibitem{Var00a} C. E. Vargas, J. G. Hirsch, and J. P. Draayer, Nucl. Phys. 
	{\bf A 673}, 219 (2000).
\bibitem{Mol95} P. Moller, J.R. Nix, and W.J. Swiatecki, Atomic Data Nucl. Data Tables 59, 185 (1995).
\bibitem{Var00b} C. E. Vargas, J. G. Hirsch, T. Beuschel, and J. P. Draayer,
	Phys. Rev. C {\bf 61}, 031301(R) (2000).
\bibitem{Var01} C. E. Vargas, J. G. Hirsch, and J. P. Draayer, Phys. Rev. C {\bf 64}, 034306 (2001).
\bibitem{Var98} C. Vargas, J. G. Hirsch, P. O. Hess, and J. P. Draayer, Phys. Rev. {\bf C 58}, 1488 (1998).
\bibitem{Var03} C. E. Vargas, J. G. Hirsch, and J. P. Draayer, Phys. Lett. {\bf B 551}, 98 (2003).
\bibitem{Rin79} P. Ring and P. Schuck. {\it The Nuclear Many-Body
        Problem}, (Springer, Berlin, 1979).
\bibitem{Duf96} M. Dufour and A. P. Zuker, Phys. Rev. {\bf C 54}, 1641 (1996).
\bibitem{Bal01} Balraj Singh, Nucl. Data Sheets 93, 243 (2001).
\bibitem{Cor08} Coral M. Baglin, Nucl. Data Sheets 109, 1103 (2008).
\bibitem{Cor02} Coral M. Baglin, Nucl. Data Sheets 96, 611 (2002).
\bibitem{Shi94} V. S. Shirley, Nucl. Data Sheets 71, 261 (1994).
\bibitem{Leh99} H. Lehmann, H. G. Borner, R. F. Casten, F. Corminboeuf, C. Doll, M. Jentschel, 
	J. Jolie and N V Zamfir, J. Phys. G: Nucl. Part. Phys. 25, 827 (1999).

\end{thebibliography}
\end{document}